\documentclass[
    ,final            
  ]
  {aipproc}

\layoutstyle{6x9}

\newcommand{\ba}{\begin{eqnarray}}
\newcommand{\ea}{\end{eqnarray}}
\begin{document}

\title{Critical-Point Structure in Finite Nuclei}

\classification{21.60Fw, 21.10Re, 05.70.Fh}
\keywords      {Quantum shape-phase transitions, critical-points, 
interacting boson model of nuclei}

\author{A. Leviatan}{
  address={Racah Institute of Physics, The Hebrew University, 
Jerusalem 91904, Israel}
}



\begin{abstract}
Properties of quantum shape-phase transitions in finite nuclei are 
considered in the framework of the interacting boson model. 
Special emphasis is paid to the dynamics at the critical-point of a 
general first-order phase transition. 
\end{abstract}

\maketitle


Phase transitions associated with a change of shape are known to occur 
in dynamical systems such as nuclei.
Recently, it has been recognized that such 
quantum shape-phase transitions are amenable to analytic descriptions 
at the critical points~\cite{iac00,iac01}. 
For nuclei these analytic benchmarks of criticality 
were obtained in the geometric framework 
of a Bohr Hamiltonian for macroscopic quadrupole shapes.
In particular, the E(5)~\cite{iac00} (X(5)~\cite{iac01}) benchmark 
is applicable to a second- (first-) order shape-phase transition between
spherical and deformed $\gamma$-unstable (axially-symmetric) nuclei. 
Empirical evidence of these benchmarks have been 
presented~\cite{casten00,casten01}. 
An important issue concerning phase transitions in real nuclei is the 
role of a finite number of nucleons. This aspect can be addressed 
in the algebraic framework of the interacting 
boson model (IBM)~\cite{ibm} which describes low-lying quadrupole collective 
states in nuclei in terms of a system of $N$ monopole ($s$) and
quadrupole ($d$) bosons representing valence nucleon pairs. 
The three dynamical symmetry limits of the model: U(5), SU(3), and O(6), 
describe the dynamics of stable nuclear shapes: spherical, axially-deformed, 
and $\gamma$-unstable deformed.
A geometric visualization of the model is obtained by 
an intrinsic energy surface defined by 
the expectation value of the Hamiltonian in the coherent (intrinsic) 
state~\cite{diep80,gino80}
\ba
\vert\,\beta,\gamma ; N \rangle &=&
(N!)^{-1/2}(b^{\dagger}_{c})^N\,\vert 0\,\rangle ~,
\label{cond}
\ea
where $b^{\dagger}_{c} = (1+\beta^2)^{-1/2}[\beta\cos\gamma\,
d^{\dagger}_{0} + \beta\sin{\gamma}\,
( d^{\dagger}_{2} + d^{\dagger}_{-2})/\sqrt{2} + s^{\dagger}\,]$.
For the general IBM Hamiltonian with one- and two-body interactions, the 
energy surface takes the form
\ba
E(\beta,\gamma) &=&
N(N-1)(1+\beta^2)^{-2}\,
\left [ a\beta^{2} - b\beta^3\cos 3\gamma + c\beta^4\right ]~.
\ea
The coefficients $a,b,c$ involve particular linear 
combinations of the Hamiltonian's parameters~\cite{kirlev85,lev87}. 
The quadrupole shape parameters in the 
intrinsic state characterize the associated equilibrium shape. 
Phase transitions for finite N can be studied 
by an IBM Hamiltonian involving terms from different dynamical 
symmetry chains~\cite{diep80}. The nature of the phase transition is 
governed by the topology of the corresponding energy surface. In a 
second-order phase transition, the energy surface is $\gamma$-independent 
and has a single minimum which changes continuously from a spherical to 
a deformed $\gamma$-unstable phase. At the critical point, $a=b=0$, and 
the energy-surface acquires a flat behaviour ($\sim \beta^4$) for small 
$\beta$. This is the situation encountered in the U(5)-O(6) phase transition. 
In a first-order phase transition the energy surface has two coexisting 
minima which become degenerate at the critical point.
The U(5)-SU(3) phase transition is a special case of this class, 
for which the barrier separating the spherical and axially-deformed minima 
is extremely small, and hence 
the critical energy surface is rather flat. 
Numerical~\cite{casten00,iaczam04} and (approximate) 
analytic~\cite{levgin03,lev05} studies within the IBM, show that 
the U(5)-O(6) [U(5)-SU(3)] critical Hamiltonians capture the essential 
features of the E(5) [X(5)] models which employ infinite square-well 
potentials. In the present contribution we consider the properties 
of a general first-order phase transition with an arbitrary barrier.
In this case the critical energy surface satisfies $b>0$ and $b^2=4ac$, 
and for $\gamma=0$ has the form
\ba
E_{cri}(\beta) &=&
c\,N(N-1)f(\beta)
\nonumber\\
f(\beta) &=& \beta^2\,(1+\beta^2)^{-2}\,(\beta - \beta_0)^2 ~.
\ea
\begin{figure}
  \includegraphics[angle=270,totalheight=7cm]{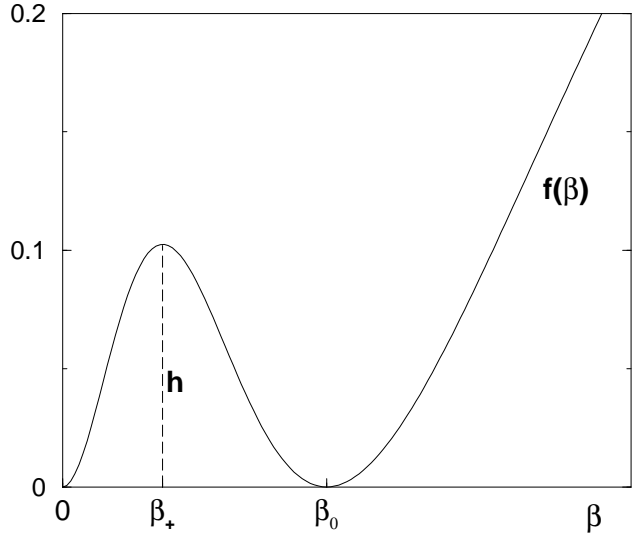}
  \caption{The IBM energy surface, Eq.~(3), at the critical point of a 
first-order phase transition. 
The position and height of the barrier are 
$\beta_{+} = \frac{-1 + \sqrt{1+\beta_{0}^2}}{\beta_0}$ and  
$h = f(\beta_{+}) = 
\frac{1}{4}\,\left ( -1 + \sqrt{1+\beta_{0}^2}\;\right )^2$ respectively.}
\end{figure}
As shown in Fig.~1, $E_{cri}(\beta)$ exhibits degenerate spherical and 
deformed minima, at $\beta=0$ and $\beta=\beta_0 =\frac{2a}{b} >0$. 
The value of $\beta_0$ determines the position $(\beta=\beta_{+})$ 
and height $(h)$ of the barrier separating the two minima in a 
manner given in the caption. 
To construct an Hamiltonian with such an energy surface, it is advantageous 
to resolve the critical Hamiltonian into intrinsic and 
collective parts~\cite{kirlev85,lev87}
\ba
H &=& H_{int} + H_c ~.
\ea
The intrinsic part of the Hamiltonian ($H_{int}$) 
is defined to have the equilibrium 
condensate $\vert \,\beta=\beta_0,\gamma=0 ; N\rangle$ as an exact 
zero-energy eigenstate and to have an energy surface as in Eq. (3).
\begin{figure}
\resizebox{7cm}{!}
{\includegraphics[angle=270,width=0.49\textwidth]{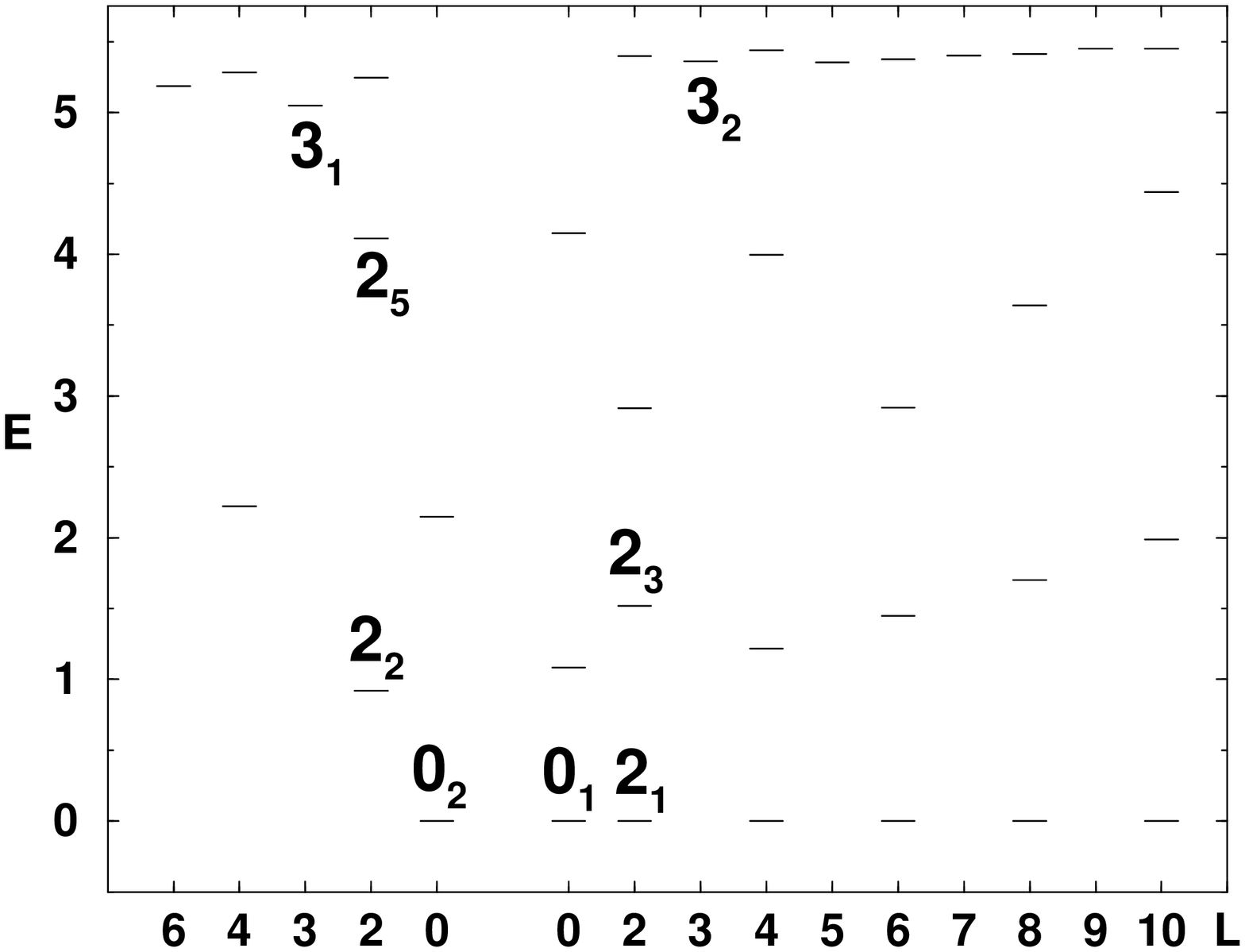}}
\resizebox{7.2cm}{!}
{\includegraphics[angle=270,width=0.49\textwidth]{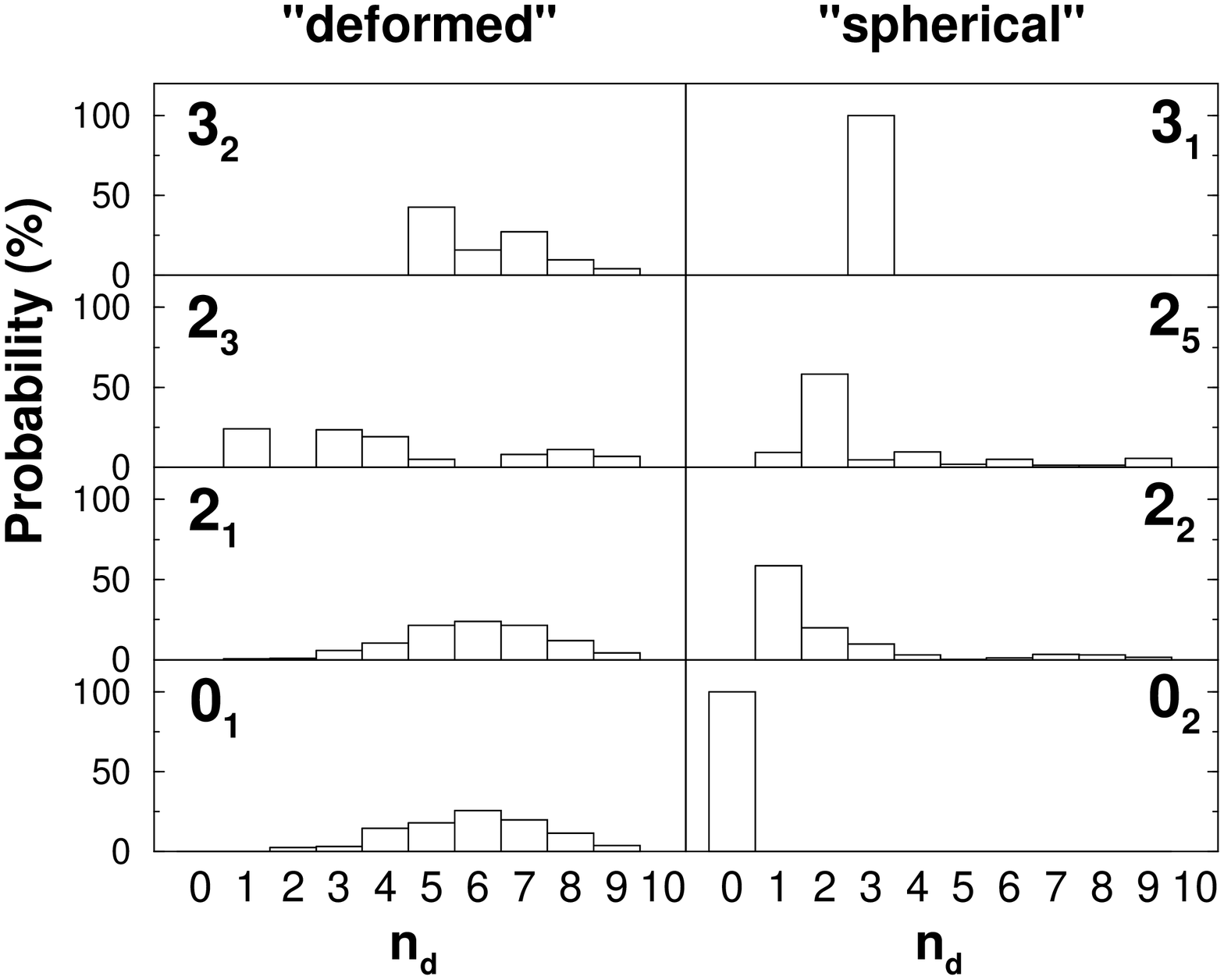}}
  \caption{Left potion: spectrum of $H_{int}$, Eq.~(5), with $h_2=0.1$, 
$\beta_0 =1.3$ and $N=10$. Right~portion: the number of $d$ bosons ($n_d$) 
probability distribution for selected eigenstates of $H_{int}$.}
\end{figure}
It can be transcribed in the form
\ba
H_{int} &=& h_{2}\,P^{\dagger}_{2}\cdot\tilde{P}_2  ~,
\ea
with 
$P^{\dagger}_{2,\mu} = 
\beta_{0}\,s^{\dagger}d^{\dagger}_{\mu} + 
\sqrt{\frac{7}{2}}\,\left( d^{\dagger} d^{\dagger}\right )^{(2)}_{\mu}$.
By construction $H_{int}$ has a set of solvable deformed eigenstates with 
energy $E=0$, which are the states 
$\vert \beta=\beta_0;N,L\rangle$
with angular momentum $ L=0,2,4,\ldots,2N$ projected from 
the intrinsic state $\vert\beta=\beta_0,\gamma=0; N\rangle$. 
It has also solvable spherical eigenstates:  
$\vert N,n_d=\tau=L=0 \rangle\equiv \vert s^N\rangle$ and 
$\vert N,n_d=\tau=3,L=3 \rangle$ with energy $E=0$ 
and $E = 3 h_2\left [\beta_{0}^2 (N-3) + 5 \right ]$ respectively.
For large $N$ the spectrum of $H_{int}$ is harmonic, involving 
5-dimensional quadrupole vibrations about the spherical minimum with frequency 
$\epsilon$, and both $\beta$ and $\gamma$ vibrations about the deformed 
minimum, with frequencies $\epsilon_{\beta}$ and $\epsilon_{\gamma}$ 
given by
\ba
\epsilon=\epsilon_{\beta}=h_{2}\,\beta_{0}^2 N \;\;\; , \;\;\;
\epsilon_{\gamma} = \frac{9}{1+\beta_{0}^2}\,\epsilon_{\beta} ~.
\ea  
For the acceptable range $0\leq\beta_0\leq 1.4$, the $\gamma$-band is 
expected to be considerably higher than the $\beta$-band. All these features 
are present in the exact spectrum of $H_{int}$ shown in Fig.~2, which 
displays a zero-energy deformed ($K=0$) ground band, degenerate with a 
spherical $(n_d=0)$ ground state. The remaining states are either 
predominantly spherical, or deformed states arranged in several excited 
$K=0$ bands below the $\gamma$ band. The coexistence of spherical and 
deformed states is evident in the right portion of Fig.~2, which shows 
the $n_d$ decomposition of wave functions of selected eigenstates of 
$H_{int}$. The ``deformed'' states show a broad $n_d$ distribution typical 
of a deformed rotor structure. The ``spherical'' states show the 
characteristic dominance of single $n_d$ components that one would expect 
for a spherical vibrator. 
In particular, as mentioned above, the solvable $L=0^{+}_2$ and $L=3^{+}_1$ 
states are pure $U(5)$ states.
\begin{figure}
\resizebox{7cm}{!}
{\includegraphics[angle=270,width=0.49\textwidth]{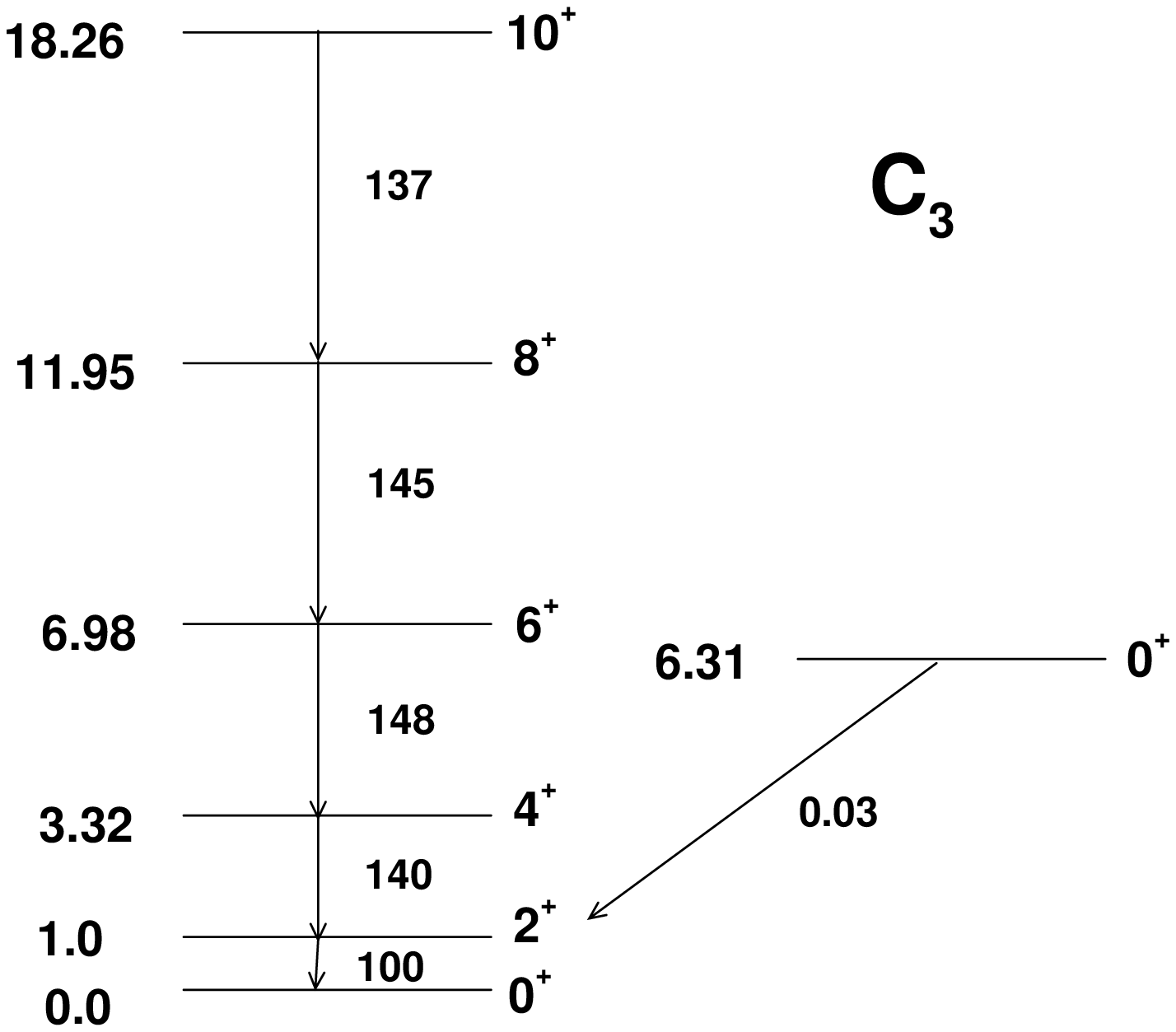}} 
\resizebox{7cm}{!}
{\includegraphics[angle=270,width=0.49\textwidth]{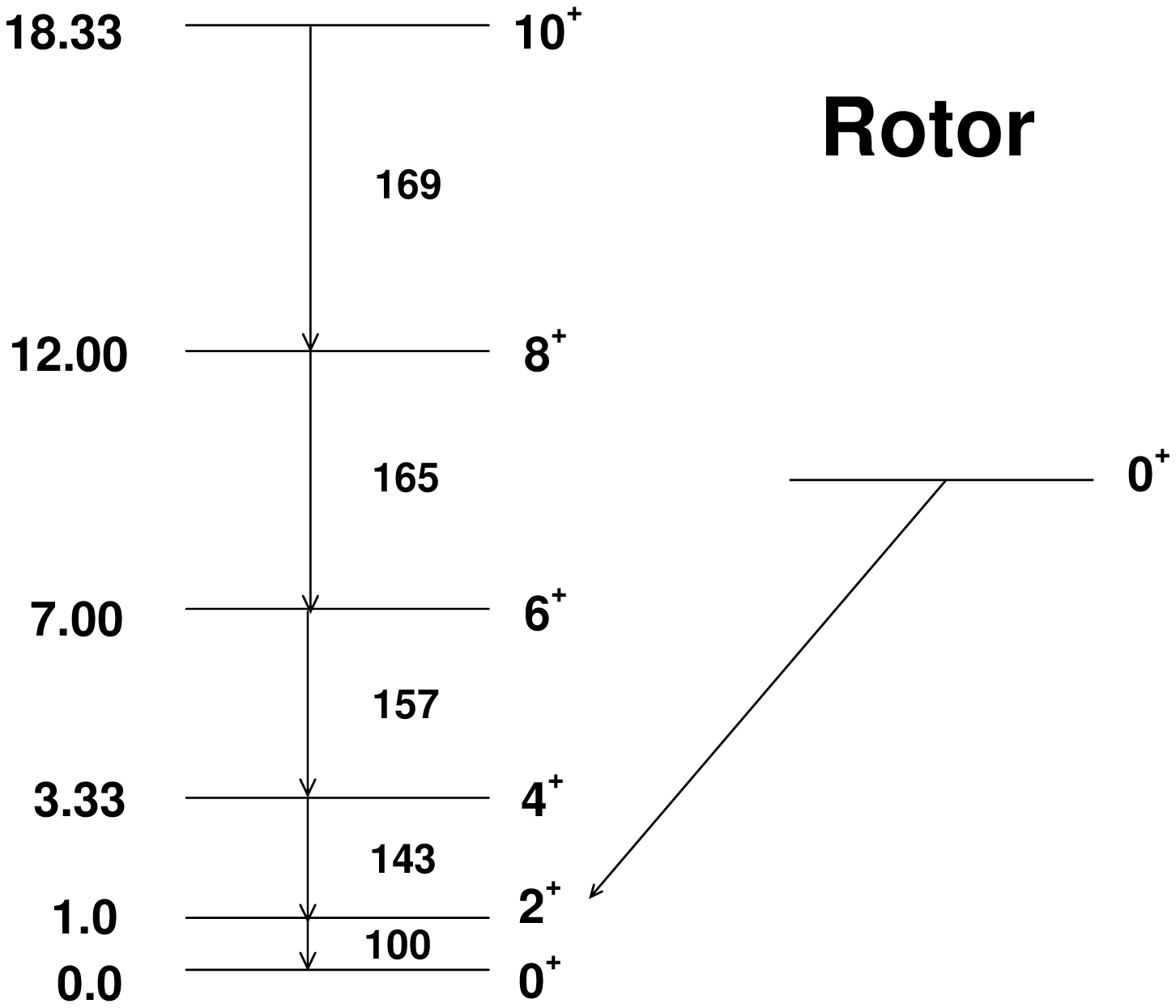}}
\end{figure}
\begin{figure}
\resizebox{7cm}{!}
{\includegraphics[angle=270,width=0.3\textwidth]{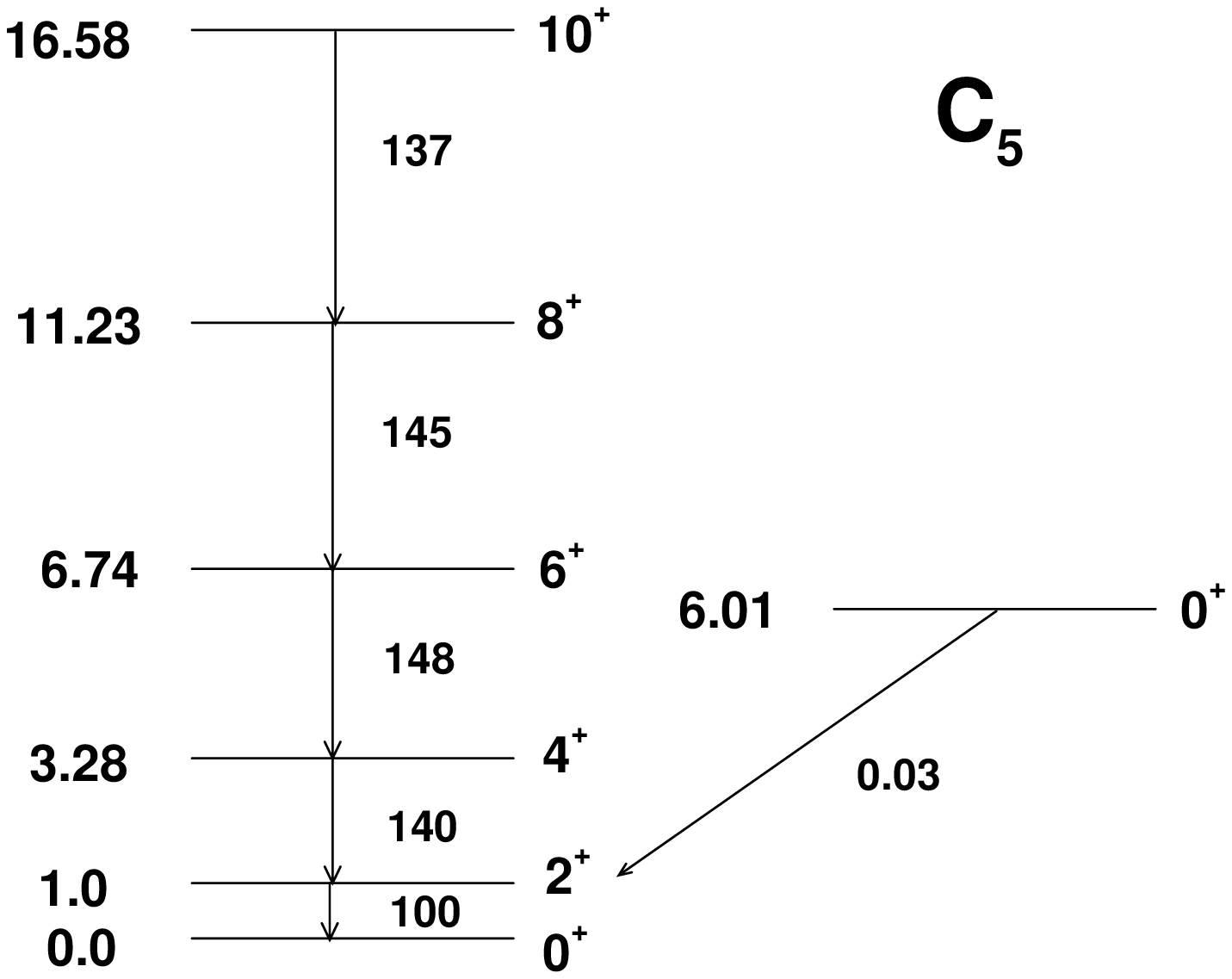}}
\end{figure}
\begin{figure}
\resizebox{7cm}{!}
{\includegraphics[angle=270,width=0.49\textwidth]{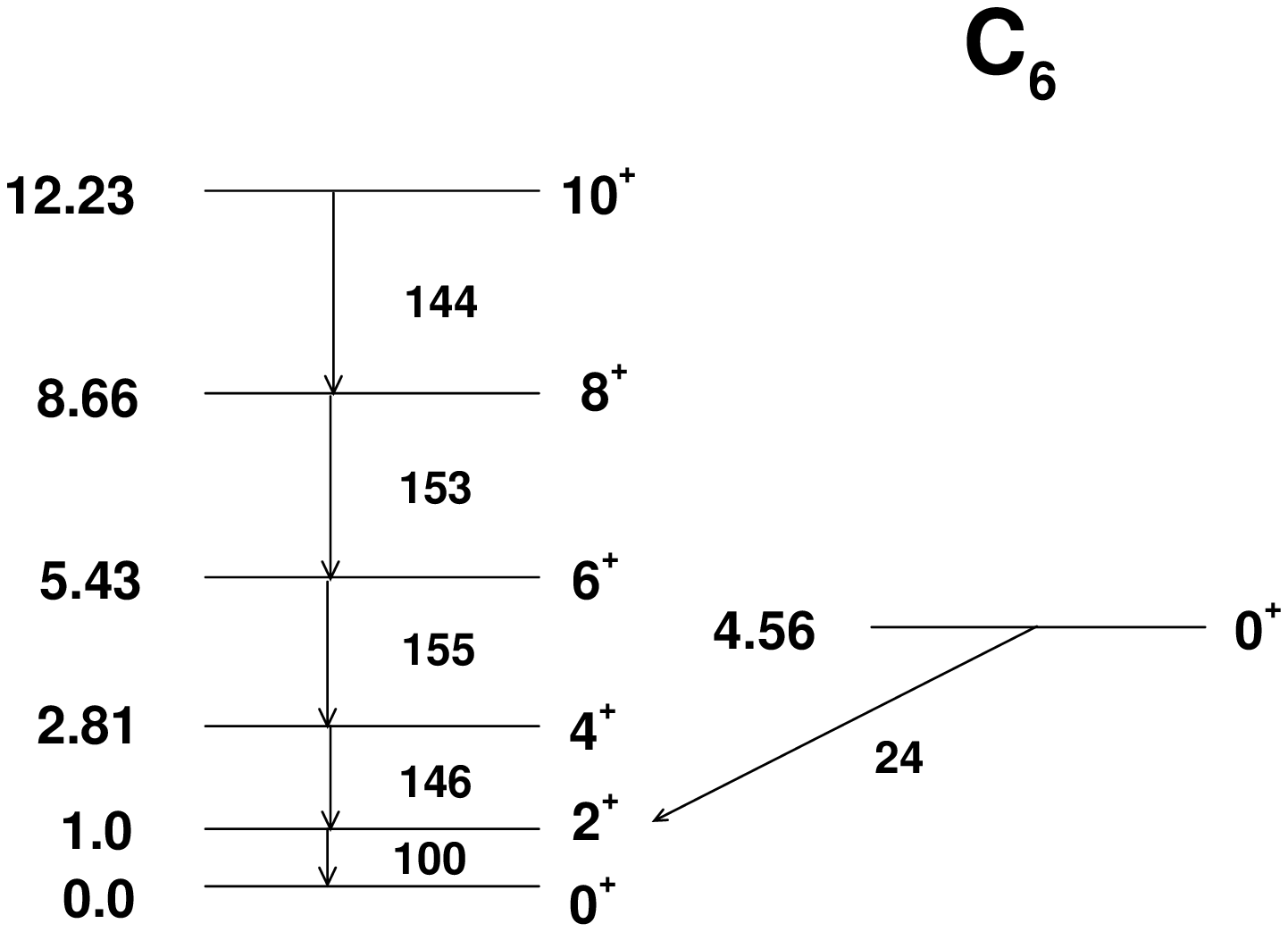}}
\resizebox{7cm}{!}
{\includegraphics[angle=270,width=0.49\textwidth]{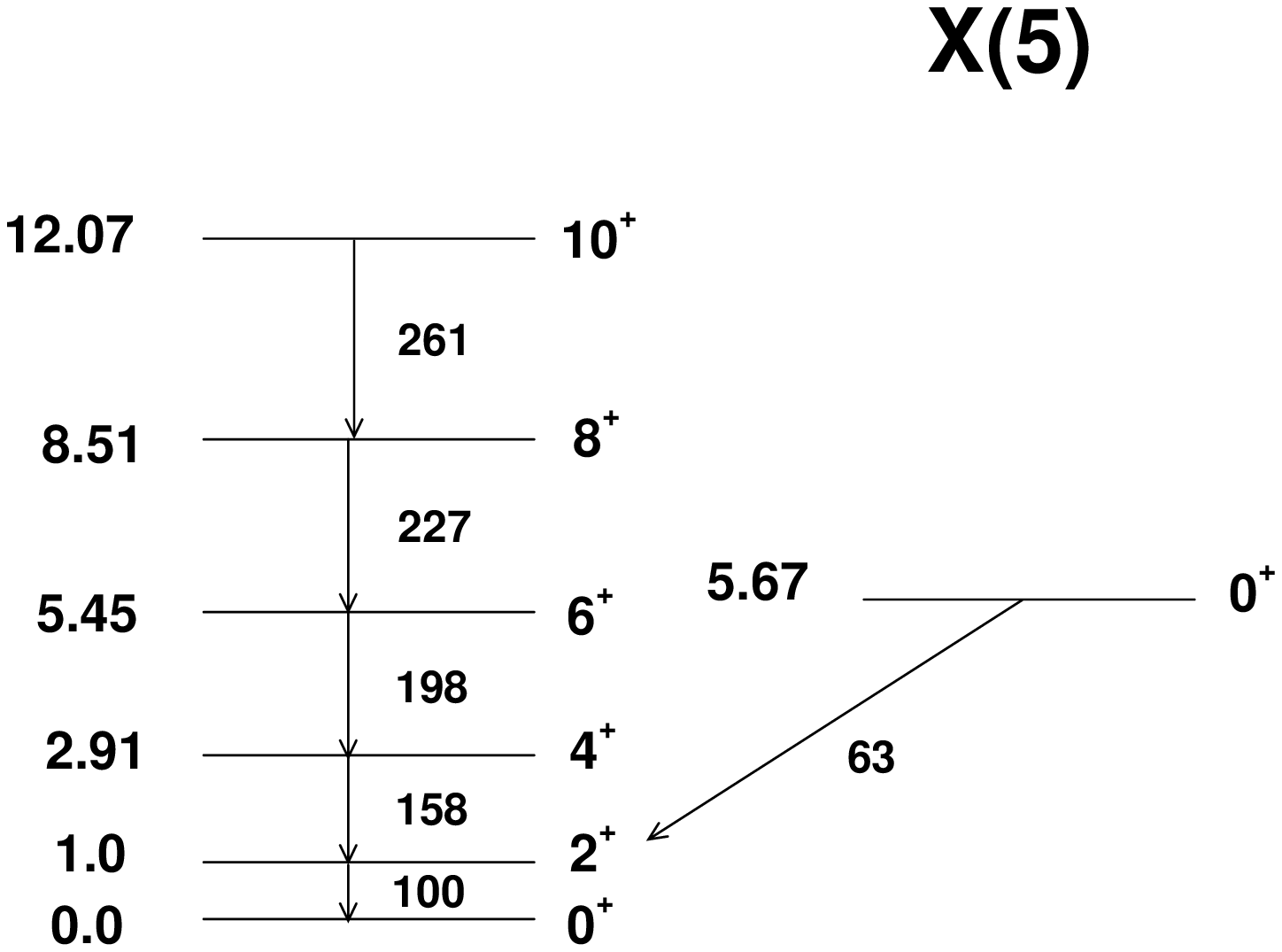}}
\caption{The spectrum of the critical Hamiltonian $H=H_{int}+H_c$, 
Eqs.~(5) and (7), with 
$\beta_0=1.3$, $N=10$, and $c_3/h_2 = 0.05$ or $c_5/h_2 = 0.1$ or 
$c_6/h_2=0.05$. The spectrum of a rigid rotor and the X(5) model is shown 
for comparison. Energies are in units of the first excited state,  
$E(2^{+}_1)-E(0^{+}_1)=100$, and $B(E2)$ values are in units of 
$B(E2;2^{+}_1\rightarrow 0^{+}_1)=100$.}
\end{figure}

The collective part ($H_c$) of the full Hamiltonian, Eq.~(4), is composed 
of kinetic terms which do not affect the shape of the energy surface. It 
can be transcribed in the form
\ba
H_{c} &=& c_3 \left [\, \hat{C}_{O(3)} - 6\hat{n}_d \,\right ]
+ c_5 \left [\, \hat{C}_{O(5)} - 4\hat{n}_d \,\right ] 
+ c_6 \left [\, \hat{C}_{\overline{O(6)}} - 5\hat{N}\,\right ] ~,
\ea
where $\hat{n}_d$ and $\hat{N}=\hat{n}_d+\hat{n}_s$ are the $d$-boson and 
total-boson number operators respectively. 
Here $\hat{C}_{G}$ denotes the quadratic Casimir operator of the 
group G as defined in \cite{lev87}. In general $H_c$ splits and mixes 
the states of $H_{int}$. Fig.~3 shows the 
effect of different rotational terms in $H_c$, 
with parameters indicated in the caption. 
For the high-barrier case considered here, ($\beta_0=1.3$, $h=0.1$), 
the calculated spectrum resembles a rigid-rotor $(E\sim a_{N}L(L+1)$) for 
the $c_3$-term, a rotor with centrifugal stretching 
$(E\sim a_{N}L(L+1) - b_{N}[L(L+1)]^2)$ for the $c_5$-term, and a 
X(5)-like spectrum for the $c_6$-term. In all cases the B(E2) values are 
close to the rigid-rotor Alaga values. This behaviour is different from 
that encountered when the barrier is low, {\it e.g.}, for the U(5)-SU(3) 
critical Hamiltonian with $\beta_0=1/2\sqrt{2}$ and $h\approx 10^{-3}$, 
where both the spectrum and E2 transitions are similar to the X(5) 
predictions. Considerable insight of the underlying 
structure at the critical point is gained by examining the 
$2\times 2$ mixing matrix between the spherical ($\vert s^N\rangle$) 
and deformed ($\beta;N,L=0\rangle$) states. The derived eigenvalues 
serve as eigenpotentials, and the corresponding eigenvectors are identified 
with the ground- and first-excited $L=0$ states. The deformed states 
$\vert \beta;N,L\rangle$ with $L>0$ are identified with excited members of 
the ground-band with energies given by the L-projected energy surface
$E_{L}^{(N)} = \langle\beta;N,L\vert H_{int} + H_{c}\vert \beta;N,L\rangle$, 
which can be evaluated in closed form 
\ba
E_{L}^{(N)}(\beta) &=& 
h_2\,(\beta-\beta_0)^2\Sigma_{2,L}^{(N)} + 
c_3\left [ L(L+1) - 6D_{1,L}^{(N)}\right ]
+ c_5\left [ -\beta^4\,S_{2,L}^{(N)} + D_{2,L}^{(N)}\right ]
\nonumber\\
&&
+\, c_6\left [ -(1+\beta^2)^2\,S_{2,L}^{(N)} +N(N-1)\right ]~.
\ea
Here
$\Sigma_{2,L}^{(N)}$, $D_{1,L}^{(N)}$, $D_{2,L}^{(N)}$ and  
$S_{2,L}^{(N)}$ denote the expectation values of $\hat{n}_s\hat{n}_d$, 
$\hat{n}_d$, $\hat{n}_d(\hat{n}_d-1)$ and 
$\hat{n}_s(\hat{n}_s-1)$ respectively in $\vert \beta;N,L\rangle$.
The value of $\beta$ in the indicated 
wave functions and energies, is chosen 
at the global minimum of the lowest eigenpotential. This procedure yields
an excellent approximation to the structure of yrast states 
at the critical point, which is then used to derive accurate finite-N 
estimates to the corresponding energies and E2 rates~\cite{lev05b}. 
The same prescription is applicable also to a first-order phase transition 
with a low-barrier \cite{lev05}.
For a second-order critical point, the determination of the effective 
$\beta$-deformation involves an $O(5)$ projection without two-level 
mixing~\cite{levgin03}.

This work was supported by the Israel Science Foundation.

\end{document}